# Critical current densities and microstructures in Rod-in-Tube and Tube Type Nb$_3$Sn strands – Present status and prospects for improvement

X Xu[1], M D Sumption[1], S Bhartiya[1], X Peng[2], and E W Collings[1]

[1] Center for Superconducting and Magnetic Materials, Department of Materials Science and Engineering, the Ohio State University, Columbus, OH 43210, U.S.A.

[2] Hyper Tech Research Incorporated, Columbus, OH 43212, U.S.A.

E-mail: xu.452@osu.edu

**Abstract**

In this work, the transport and magnetization properties of distributed-barrier Rod-in-Tube (RIT) strands and Tube Type strands are studied. While Tube Type strands had smaller magnetizations and thus better stabilities in the low field region, their 12 T non-Cu $J_c$s were somewhat smaller than those of the RIT strands. Microstructures were investigated in order to find out the reasons for the difference in non-Cu $J_c$ values. Their grain size and stoichiometry were found to be comparable, leading to similar layer $J_c$s. Accordingly it was determined that the lower A15 area fraction rather than the quality of A15 layer was the cause of the discrepancy in non-Cu $J_c$. Subsequently, the area utilizations of subelements were investigated. While for a RIT strand the fine grain (FG) A15 area occupies ~60% of a subelement, for a Tube Type strand it is no more than 40%. Further analysis indicates that the low FG area fraction in a Tube Type strand is attributed to its much larger unreacted Nb area fraction. Finally, a simple change in strand architecture is proposed to reduce the unreacted Nb area fraction.

## 1. Introduction

Nb$_3$Sn is the present conductor of choice for the construction of high field magnets (above 8 T), and has significant application prospects in high energy physics (HEP) and fusion devices. YBCO coated conductor and Bi:2212 are attractive, however a number of issues, including high cost, have so far limited the application of these conductors for large scale magnets. On the other hand, high performance Nb$_3$Sn conductors have an appropriate mix of superconducting properties for the enabling of magnets in above 15 T range. In the past two decades the 4.2 K, 12 T non-Cu critical current density ($J_c$) values have doubled [1]. The present record of above 3000 A/mm$^2$ (at 4.2 K, 12 T) is held by the distributed barrier internal-Sn (or Rod-in-Tube, alternatively Restacked-Rod- Process, RRP) strands, and are thus one of the favored advanced conductors for near term HEP machine development and machine upgrades. The goal of internal-Sn strand processing is to achieve high $J_c$ in association with a small subelement diameter, $d_{eff}$, since a small $J_c d_{eff}$ product is necessary for the low field stability essential for most applications of Nb$_3$Sn [2, 3, 4]. A small $J_c d_{eff}$ product is also very important in multi-pole magnets for improving field quality at the bore because reduction in persistent current magnetization at low fields calls for a reduction in $J_c d_{eff}$ product [5]. In 1999 the U.S. Conductor Development Program set a goal for the HEP conductors, which includes but not limited to a 12 T non-Cu $J_c$ of 3000 A/mm$^2$ and $d_{eff}$ of less than 40 μm [6]. Reductions in $d_{eff}$ can be realized by increasing the number of subelements in a strand. Oxford Superconducting Technology (OST) has successfully produced 217-subelement RIT strands with $d_{eff}$s as low as 43 μm accompanied by 12 T non-Cu $J_c$s above 2600 A/mm$^2$ [7]. Nevertheless, serious bonding issues [8] and metalworking difficulty are encountered during the drawing of a RIT strand to $d_{eff}$s < 40 μm. The RIT subelement is an annulus of some 1000 Cu-clad Nb rods (Nb "filaments") surrounding a Sn core. When the multi-subelement precursor strand is drawn down to a final size of some 0.7-1 mm the annulus evolves into a Nb/Cu microcomposite [9] subjected to Hall-Petch (H-P) hardening which sets in as the thickness of the interfilamentary Cu drops below 10 µm and H-P plus interface strengthening as the Cu thickness enters the nanometer range. Although advantage has been taken of these hardening mechanisms in the development of high strength high conductivity Nb/Cu microcomposites [10] for magnet applications [11, 12], they are a serious disadvantage to internal-Sn RIT strand fabrication. For a commercial 1000-filament 60 µm subelement we estimate the "geometrically ideal" Nb-Nb separation to be around 0.15 µm on average which places it firmly in the H-P plus surface-hardening regime. To eliminate such hardening during drawing to fine subelement diameters it would in principle be possible to use fewer but larger diameter filaments the ultimate limit to which is the idea of encircling a Cu-clad Sn core with a Nb tube. Two strand designs which, in effect, do this are (i) powder-in-tube (PIT) and (ii) Tube-Type strands. Powder-in-tube strands have lower subelement diameters (higher subelement counts) than RIT strands, but the need for very fine powders results in a much higher cost. The Tube Type process, on the other hand, allows for high filament counts with a very simple geometry, and thus has in principle the potential for low cost [13]. In this method the subelement consists of a Cu-clad Sn rod inserted into a Nb tube. This simple subelement construction is amenable to heavy reduction in size during wire drawing. At present, Hyper Tech Research Inc (HTR) has successfully produced 1387-subelement Tube Type strands, reducing the subelement diameter to 12 µm [14]. In spite of its advantages of small $d_{eff}$ and stability the present Tube Type strand has a lower non-Cu $J_c$ than its RIT counterpart [15]. This report addresses the cause of the discrepancy and offers a solution. In so doing the comparative properties of RIT and Tube Type strands are discussed in detail in terms of magnetization, transport properties, chemistry, grain sizes, and A15 area utilization.

## 2. Experiment

## 2.1. Strands

Two types of Nb$_3$Sn strands were used in this work: the RIT type and Tube Type, all manufactured by HTR.

Four RIT strands were investigated in this work. Their billet designs are described in Table 1. The billet designs of the three unreacted Tube Type strands are shown in Table 2, and BSE images of the unreacted strands can be found in [16].

## 2.2. Heat Treatment and Measurement

Both types of strand were prepared for measurement in two ways: (i) long samples were wound and heat treated on ITER barrels for transport $J_c$ measurement, (ii) short, straight samples were prepared for magnetic measurements and scanning electron microscopy (SEM), to enable superconducting properties to be correlated with the microstructures.

The heat treatment schedules for RIT strands are given in Table 3 (a). For RIT strands, preheat treatments (indicated by "Pre") were applied prior to the reaction heat treatments, including 48 hours at 210 ˚C (ramp rate: 10 ˚C/h) and 48 hours at 400 ˚C (ramp rate 25 ˚C/h). They were used for the Sn to diffuse into Cu. Then the reaction heat treatments (ramp rate: 50 ˚C/h) were applied.

For Tube Type strands there were no pre-heat treatments. Each Tube Type strand underwent several reaction heat treatments (including some two-step ones) so that the optimum heat treatment schedules with respect to the non-Cu $J_c$ can be found. These heat treatments are given in Table 3 (b). All the strands were furnace-cooled after the heat treatments.

### 2.2.1. Transport Measurements

Transport $J_c$ was measured on ITER type barrels machined from Ti-6Al-4V alloy and fitted with Cu end-rings. It is well known that the strain state of the Nb$_3$Sn layer has an influence on the $J_c$ [17, 18]. Here we use a modified ITER Barrel arrangement, in which both the Ti-6Al-4V and the Cu end rings were threaded into matching helical grooves, to minimize the strand motion. This configuration was used for both RIT and Tube Type strands during $J_c$ measurement, leaving the thermal compression the only possible strain source. It has been shown by [18] that the pre-compression strains for RIT and Tube Type strands are at the same level. Unreacted pieces of strand about 1.5 m in length with both ends sealed with a blowtorch, wound around the groove in the barrel and fixed to the Cu end rings with stainless steel screws, were heat treated in a furnace under flowing argon. In preparation for measurement the ends of the strands were soldered down to the Cu end rings. The barrels were then mounted on the current-carrying probe, and voltage taps were soldered on at a separation (gauge length) of 50 cm. A thin layer of thermally conductive blue stycast was painted on the barrel to prevent the strand motion. The measurements were performed in liquid helium at 4.2 K in transverse fields of from 0 T to 14 T; the transport $J_c$ was determined at a voltage criterion of 0.1 µV/cm. A more detailed description of the transport measurement can be found in [16].

*2.2.2. Magnetic Measurements and Electron Microscopy*

Straight samples about 25 cm in length were prepared for magnetic measurement and SEM/EDS. Before heat treatment they were torched on their two ends to seal them against subsequent Sn leakage and then encapsulated in quartz tubes under 200 torr (or 1.33 mbar) of Argon for the heat treatment. The heat treatment schedules were the same with those applied to the transport measurement strands.

Magnetic measurements were performed using the Vibrating Sample Magnetometer (VSM) function of a Quantum Design Model 6000 "physical property measuring system"

(PPMS) on strand pieces about 4-5 mm length removed from the center of the heat treated short sample. $M$-$\mu_0H$ loop measurements were performed at 4.2 K on samples oriented perpendicular to applied fields of ± 14 T at a ramp rate of 13 mT/s.

For BSE/EDS studies the reacted samples were mounted in conductive Bakelite powder and polished to 0.05 μm. An FEI QUANTA 200 SEM with EDS attachment was used to perform the SEM studies. While in the EDS mode, 20 kV was used for the accelerating voltage. For each spot data were collected for 200 seconds.

SEM images of fracture surfaces of samples were used to determine A15 grain sizes. The SEM images for this analysis were taken on a Sirion field emission SEM which has a spatial resolution of 1-3 nm in ultra-high resolution mode.

## 3. Results

### 3.1. Transport $J_c$ of RIT and Tube Type strands

Several heat treatments were performed for each Tube Type strands in search of the optimum Heat treatment schedule. The transport 4.2 K non-Cu $J_c$ values versus applied fields are shown in figure 1. From figure 1 it is determined that the best heat treatments are: 625x500 for T1505, 625x300 for T1628, and 615x480 for T1489. It is clear that the lower temperature and longer time heat treatments work better for the Tube Type strands. It is interesting to note that a similar conclusion was reached for PIT strands whose best HT was 625x320h [19].

Figure 2 shows the transport 4.2 K non-Cu $J_c$ values of RIT and optimally heat treated Tube Type strands from 11 T to 14 T. The 4.2 K, 12 T non-Cu $J_c$s of the Tube Type samples, which range from 1800 to 2500 A/mm$^2$, are lower than those of the RIT strands, which are within 2700~3500 A/mm$^2$.

The irreversibility field ($\mu_0H_{irr}$) values were derived by extrapolating the intermediate field Kramer plots into zero, and are presented in Table 4. The extrapolated $\mu_0H_{irr}$s of the Tube Type strands are around 25~26 T; those of the RIT strands were generally about 2 T lower (with the exception of EG36Ti-61Re-650x60, in which the Ti doping has increased the extrapolated $\mu_0H_{irr}$s by about 3 T).

*3.2. Magnetic studies of RIT and Tube Type strands*

The magnetization versus applied magnetic field ($M$-$\mu_0H$) loops for RIT strands and Tube Type strands are shown in figure 3. With zero-field peaks at around $2 \times 10^5$ A/m compared to about $5 \times 10^5$ A/m the magnetizations of the Tube Type strands are much smaller than those of the RIT strand. For this reason the Tube Type strands are free from low field flux jumping while the RIT strands exhibit copious flux jumping in fields below about 2.5 T.

*3.3. Stoichiometry and Grain Size Analysis*

*3.3.1. Stoichiometry*

Figure 4 (a) shows the fracture SEM image of a subelement of EG36-91Re-650x50 including the core/A15 boundary, with respect to which the radial distances of the EDS spots (indicated by the white spots in figure 4 b) were measured. These spots were manually located at the centers of the original filaments to avoid interaction with the Cu islands. The A15 areas of Tube Type samples are mainly comprised of two parts: the fine grain (FG) area and the coarse grain (CG) area, as can be seen in figure 4 (c) which shows the fracture SEM image of a subelement of T1489-615x480. For Tube Type strands the radial distances of EDS spots were measured from the CG/FG boundary. Figure 4 (d) shows the Sn content profiles of these two strands.

From figure 4 it can be seen that within experimental error the Sn concentrations in Tube type and RIT strands are more or less the same (note that for RIT strands the EDS spots were located at the centers of the original filaments which could result in a slight underestimation of the Sn content because of a Sn content gradient within this region [20]). The average Sn concentrations, which were calculated by averaging the Sn concentrations of different positions in the FG layer, of EG36-91Re-650x50 and T1489-615x480 are 22.7 at.% and 23.0 at.%, respectively.

*3.3.2. Grain size*

The fracture SEM images of some Tube Type and RIT strands are shown in figure 5. The Feret grain sizes (see, for example, [21]) of the fine grains of T1489-615x480, EG36Ti-61Re-650x60, and EG36-91Re-650x50 are respectively 115, 120, and 110 nm, the differences among which are not significant. The Sn contents and grain sizes are essentially similar leading to the conclusion that the $J_c$s of the fine grain layers should also be at the same level. Thus the area fractions will be investigated below.

*3.4. Area utilization analysis*

As depicted in the BSE and fracture SEM images in figure 6, subelements of Tube Type (figure 6 a-c) are composed of four regions: core, CG, FG, and unreacted Nb, while those of RIT strands (figure 6 d-f) are comprised of the core, A15, and unreacted Nb regions -- although there are a large number of disconnected phases at the core/A15 boundary, this region carries no current and in this study is regarded as part of the core region.

Area fractions of these parts were measured and shown in Table 5, along with their layer $J_c$s. For RIT strands the layer $J_c$s were calculated by dividing the non-Cu $J_c$s by the A15 area fractions. For Tube Type strands, to avoid an overestimation of the FG layer $J_c$, the CG area is assumed to carry 10% of the current density of FG area. This comes from the ratio of CG to FG grain sizes. Some studies [22, 23] showed that pinning force $F_p$ ($=J_c \times \mu_0 H$) is inversely proportional to the grain size, whereas others suggested that $F_p$ varied faster [24] or slower [25] than linearly with the reciprocal of grain size. In this study the linear dependence relation is adopted. The CG size for our samples is in the 1 µm range, and the FG of about 0.1 µm. This leads to a factor of 1/10 multiplier for the CG contribution to overall layer $J_c$. Thus, the layer $J_c$ of Tube Type strands was calculated by dividing the non-Cu $J_c$s by the sum of FG area fractions and 1/10 of the CG area fractions. If, on the other hand, the CG area is assumed to carry no current, the calculated layer $J_c$s of Tube Type strands are nearly 4% higher than the results in Table 5. From Table 5, the HP31-61Re-650x80h has a lower 12 T layer $J_c$ compared with the other two RIT strands. This is probably in consequence of its lower Sn/Cu ratio design within the subelement (see Table 1) and the associated lower irreversibility field (as seen in Table 4). Furthermore, it is important to note that while the non-Cu $J_c$ of RIT is 40% higher than that of Tube Type, the FG-layer properties are comparable: the grain sizes, the Sn concentrations, and the layer $J_c$s are roughly the same, respectively, within the scatter of the data. What then is responsible for Tube Type's lower non-Cu $J_c$? The answer lies in the difference between the FG area fractions. In the RIT strand fine grains occupy 61 % of the subelement area as compared to the 40 % FG occupancy of the Tube Type subelement. We will go on to show that with equal FG area fractions the Tube Type's non-Cu $J_c$ could be similar to that of RIT.

## 4. Discussion

Considering their longer diffusion length, one might think that tubular strands have somewhat lower non-Cu $J_c$s, assuming they have lower Sn contents or larger Sn concentration gradients across the FG area. However, the EDS results in this work show that in reality the Sn contents in Tube Type strands are more or less at the same level with the best RIT strands.

According to Lee *et al.*'s work in the RRP strands produced by OST [26], the average Sn content in the RRP2 strand of the very highest $J_c$ (12 T) is 23 at.%, which is consistent with the results in this paper. Improving the reaction temperature can push the average Sn content to 24.0 at.%, but the average grain size is also increased remarkably [26].

Since the layer $J_c$s are the same within scatter, if the FG area fraction of the Tube Type strand is expanded to 61.3% (the average for the RIT strands) while its FG layer $J_c$ is maintained, the Tube Type strand will end up with a similar 12 T non-Cu $J_c$ with that of the RIT strand (about 3300 A/mm$^2$). What are the prospects for increasing the Tube Type's FG area fraction in order to achieve this improvement? To answer this question we revisit the architectures of the strands as summarized in Table 5, and some of the factors that influence them. The large CG grains of Tube Type can only carry a relatively small supercurrent, so it would be useful if the CG area fraction could be decreased provided the FG fraction and its properties increased at the same time. The authors have studied these issues in detail and showed that increasing the Cu/Sn ratio discourages CG growth and thus improves FG area fraction [27]. Next, Table 4 emphasizes that the greatest opportunity for FG area increase lies in the unreacted Nb region of the strand. The unreacted Nb area takes about 30% of the whole subelement in Tube Type strands compared to only 5% in RIT strands.

For the normal Tube Type strands, reducing Nb/Sn ratio in the starting strand and increasing the reaction time would certainly decrease the unreacted Nb area fraction. However, continued growth of the reaction layer carries with it the risk that Sn may diffuse out of the subelement and poison the surrounding Cu stabilizer. With hexagonal subelements it is frequently noted that the middle part of a side may become penetrated during the reaction HT while a significant fraction of the Nb may remain unreacted in the corners. For example, in a

T1489-615x480h subelement, even though the unreacted Nb area fraction still remains at 26.5%, sufficient Sn has leaked out to form a thin external layer of $Nb_3Sn$.

This kind of Sn leakage can be suppressed or averted by changing the subelement design. SEM/BSE images of Tube Type strand show the FG layer to be cylindrical and uniform in thickness. Sited within a hexagonal subelement this cylinder can touch the inside of the Nb layer while remaining far from its corners. So a way to increase the FG area fraction and reduce the unreacted Nb fraction is to begin with a round, instead of hexagonal, subelement and finish by reacting the Nb to just (but safely) below the penetration threshold.

## 5. Conclusions

This report has made a detailed examination of the relative properties of RIT and Tube Type strand noting the following conclusions. The average Sn concentrations in the FG areas are the same, ~23 ± 0.3%. The average grain sizes are the same ~ 111 ± 8 nm. For the above reasons the FG layer $J_c$s are the same ± 5%. The onset of H-P hardening makes it difficult to draw RIT strand to $d_{eff}$s below 60-30 μm. No such problem exists for Tube Type strand which as a result offers low persistent current magnetization and stability against low field flux jumping. In spite of similar layer $J_c$s, the 12 T non-Cu $J_c$ of the present Tube Type strand is less than that of the RIT, Table 5, the reason being that the Tube Type's Nb layer is not fully reacted, especially at the hexagonal corners. In a planned modification of the strand design, in which the starting billet is assembled from round rather than hexagonal subelements thereby allowing the FG area to expand almost to the edge of the Nb cylinder, we expect the Tube Type's non-Cu $J_c$ to be equal to or possibly greater than that of the RIT strand.

**Acknowledgements**

This work was funded by the US Department of Energy, Division of High Energy Physics, Grant No. DE-FG02-95ER40900 and a DOE Contract Number DE-SC0001558.

# LIST OF TABLES

Table 1. Strand Specifications of RIT Strands.

| Strand name | HP31-61Re | EG36Ti-61Re | EG36-91Re | EG36-127Re |
|---|---|---|---|---|
| Strand diameter, mm | 0.7 | 0.7 | 0.7 | 0.7 |
| No. of Subelement | 54/61[*] | 54/61 | 78/91 | 102/127 |
| Non-Cu area fraction | 52.5% | 52.5% | 49.5% | 49% |
| Tin core composition | Pure Sn | Sn-1.2% Ti | Pure Sn | Pure Sn |
| Filament composition | Nb-7.5wt%Ta | Nb-7.5wt%Ta | Nb-7.5wt%Ta | Nb-7.5wt%Ta |
| Barrier composition | Pure Nb | Pure Nb | Pure Nb | Pure Nb |
| Sn area fraction | 25.2 | 28.1 | 28.1 | 28.1 |
| Nb area fraction | 57.8 | 58.2 | 58.2 | 58.2 |
| Cu area fraction | 16.9 | 13.7 | 13.7 | 13.7 |

*54 is the number of subelement of $Nn_3Sn$, and 61 is the total number that includes Cu subelements. Similar notification will be used for all other strands in this paper.

Table 2. Strand Specifications of Tube Type Strands.

| Strand name | T1505 | T1628 | T1489 |
|---|---|---|---|
| Strand diameter, mm | 0.7 | 0.7 | 0.7 |
| No. of Subelement | 192/217 | 192/217 | 246/271 |
| Non-Cu area fraction | 46.5% | 46.5% | 46.5% |
| Tin core composition | Pure Sn | Pure Sn | Pure Sn |
| Filament composition | Nb-7.5wt%Ta | Nb-7.5wt%Ta | Nb-7.5wt%Ta |
| Sn area fraction | 21.9 | 21.8 | 22.8 |
| Nb area fraction | 73.7 | 74.2 | 72.6 |
| Cu area fraction | 4.4 | 4.0 | 4.6 |

Table 3. Heat treatment schedules for the strands in this work.

(a) RIT strands

| Strand name | HP31-61Re | EG36Ti-61Re | EG36-91Re | EG36- 127Re |
|---|---|---|---|---|
| Heat treatment, ˚C x h | Pre + 650 x 80 | Pre + 650 x 60 | Pre + 650 x 50 | Pre + 650 x 40 |

(b) Tube Type strands

| Strand name | T1505 | T1628 | T1489 |
|---|---|---|---|
| Heat treatment, ˚C x h | 625x500, 635x144,635x200, 650x72,650x100, 575x48+635x300 | 625x300, 635x250, 650x200, 575x48+635x300 | 615x480, 625x300, 615x250+650x70, 575x48+625x550, 650x72, 650x100, 650x120 |

Table 4. The extrapolated $\mu_0H_{irr}$ values for RIT and Tube Type strands.

| Sample name | Rod-in-Tube strands | | | | Tube Type strands | | |
|---|---|---|---|---|---|---|---|
| | HP31-61Re | EG36Ti-61Re | EG36-91Re | EG36-127Re | T1505 | T1628 | T1489 |
| Heat Treatment, °C x h | 650x80 | 650x60 | 650x50 | 650x40 | 625x500 | 625x300 | 615x480 |
| Extrapolated $\mu_0H_{irr}$, T | 23.5 | 26.8 | 24.2 | 23.0 | 26.1 | 25.7 | 25.9 |

Table 5. The area fractions and $J_c$ values of the RIT and Tube Type strands.

| | Heat Treatment, °C x h | Hole area, % | CG area, % | FG area, % | Unreacted Nb area, % | 4.2 K, 12 T Non-Cu $J_c$, A/mm$^2$ | 4.2 K, 12 T layer $J_c$, A/mm$^2$ |
|---|---|---|---|---|---|---|---|
| | | | Rod in Tube Strands | | | | |
| HP31-61Re | 650x80 | 35.3 | - | 61.5 | 3.2 | 3050 | 4960 |
| EG36Ti-61Re | 650x60 | 34.1 | - | 61.0 | 4.9 | 3320 | 5450 |
| EG36-91Re | 650x50 | 33.8 | - | 60.3 | 5.9 | 3480 | 5760 |
| | | | Tube Type Strands | | | | |
| T1505 | 625x500 | 16.2 | 15.9 | 41.8 | 26.1 | 2440 | 5630 |
| T1628 | 625x300 | 17.2 | 15.1 | 37.9 | 29.8 | 2150 | 5460 |
| T1489 | 615x480 | 20.4 | 14.0 | 38.9 | 26.7 | 2230 | 5530 |

# LIST OF FIGURES

**Figure 1:** The transport 4.2 K non-Cu $J_c$ values vs. applied magnetic field for the heat-treated Tube Type strands: (a) T1505, (b) T1628 and (c) T1489.

**Figure 2:** The transport non-Cu $J_c$ values (4.2 K) vs. applied magnetic field for (a) RIT and (b) Tube Type strands.

**Figure 3:** Loops of Magnetization vs. applied magnetic field for (a) RIT and (b) Tube Type strands.

**Figure 4:** (a) The fracture SEM image of a subelement of EG36-91Re-650x50 showing the core and A15 regions; (b) The BSE/SEM image of EG36-91Re-650x50, with the white dots indicating the spots where EDS were performed; (c) the fracture SEM image of a subelement of T1489-615x480 showing the CG and FG A15 regions; (d) the Sn concentration profiles of EG36-91Re-650x50 and T1489-615x480h strands.

**Figure 5:** Fracture SEM images of (a) FG and CG area of T1489-615x480, (b) FG area of T1489-615x480, (c) A15 area of EG36Ti-61Re-650x60, (d) A15 area of EG36-91Re-650x50.

**Figure 6:** SEM images of Tube Type and RIT strands: (a) BSE image of T1489-615x480, (b) fracture SEM of T1505-625x500, (c) fracture SEM of T1505-625x500 showing the core/A15 interface (d) BSE image of HP31-61Re-650x80, (e) fracture SEM of EG36Ti-61Re-650x60, (f) fracture SEM of EG36Ti-61Re-650x60 showing the core/A15 interface.

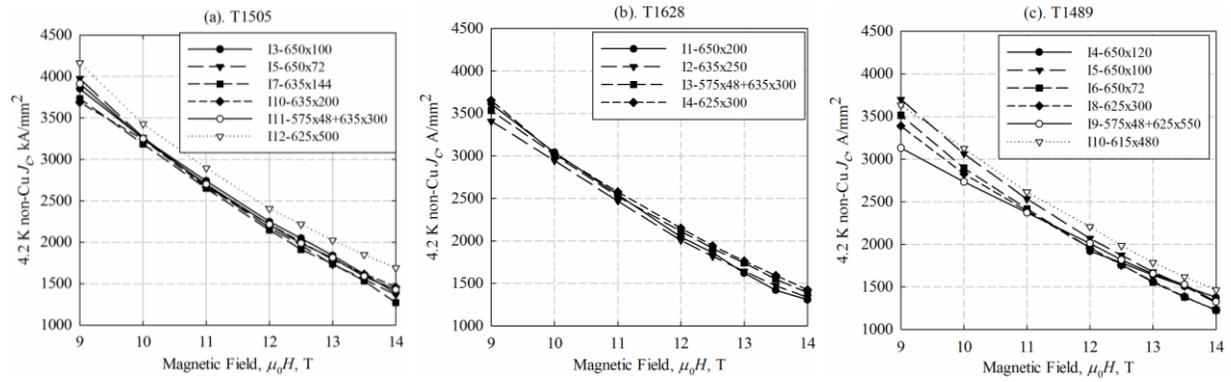

Figure 1. The transport 4.2 K non-Cu $J_c$ values vs. applied magnetic field for the heat-treated Tube Type strands: (a) T1505, (b) T1628 and (c) T1489.

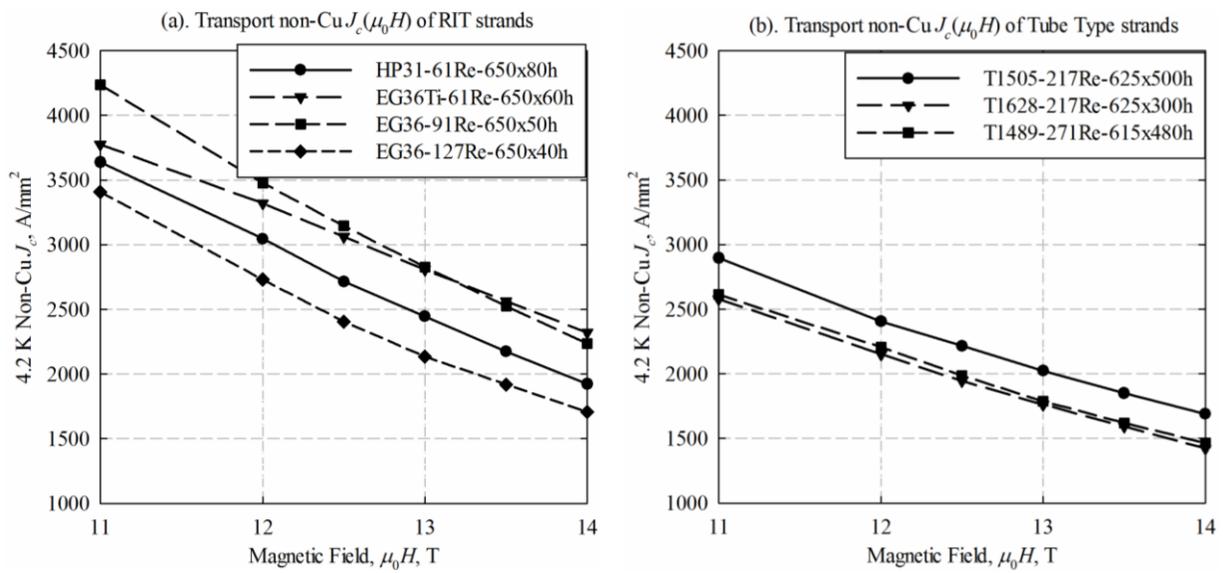

Figure 2. The transport non-Cu $J_c$ values (4.2 K) vs. applied magnetic field for (a) RIT and (b) Tube Type strands.

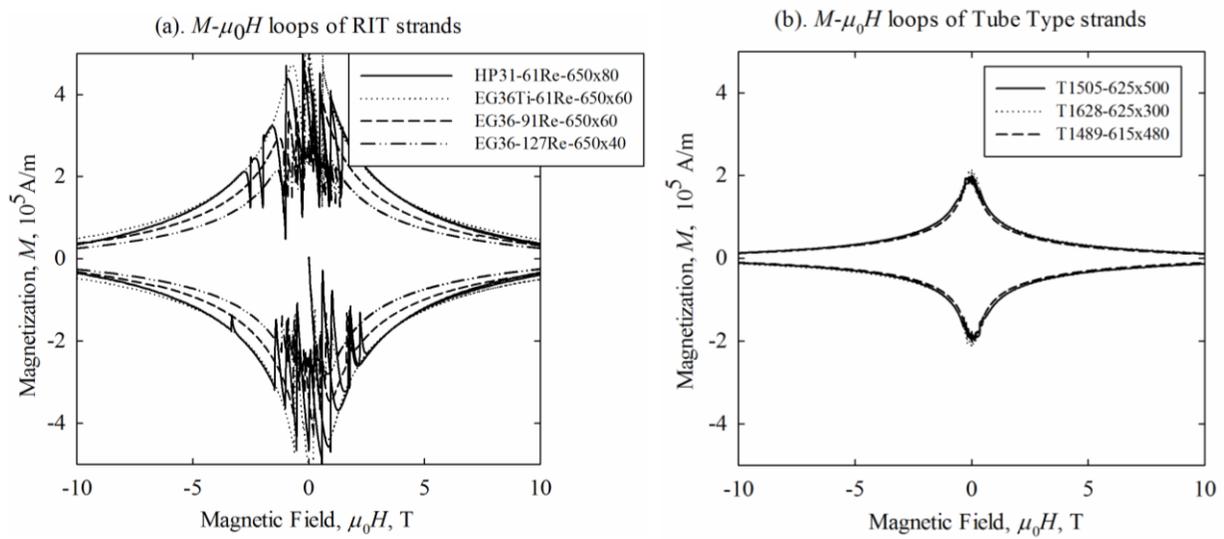

Figure 3. Loops of Magnetization vs. applied magnetic field for (a) RIT and (b) Tube Type strands.

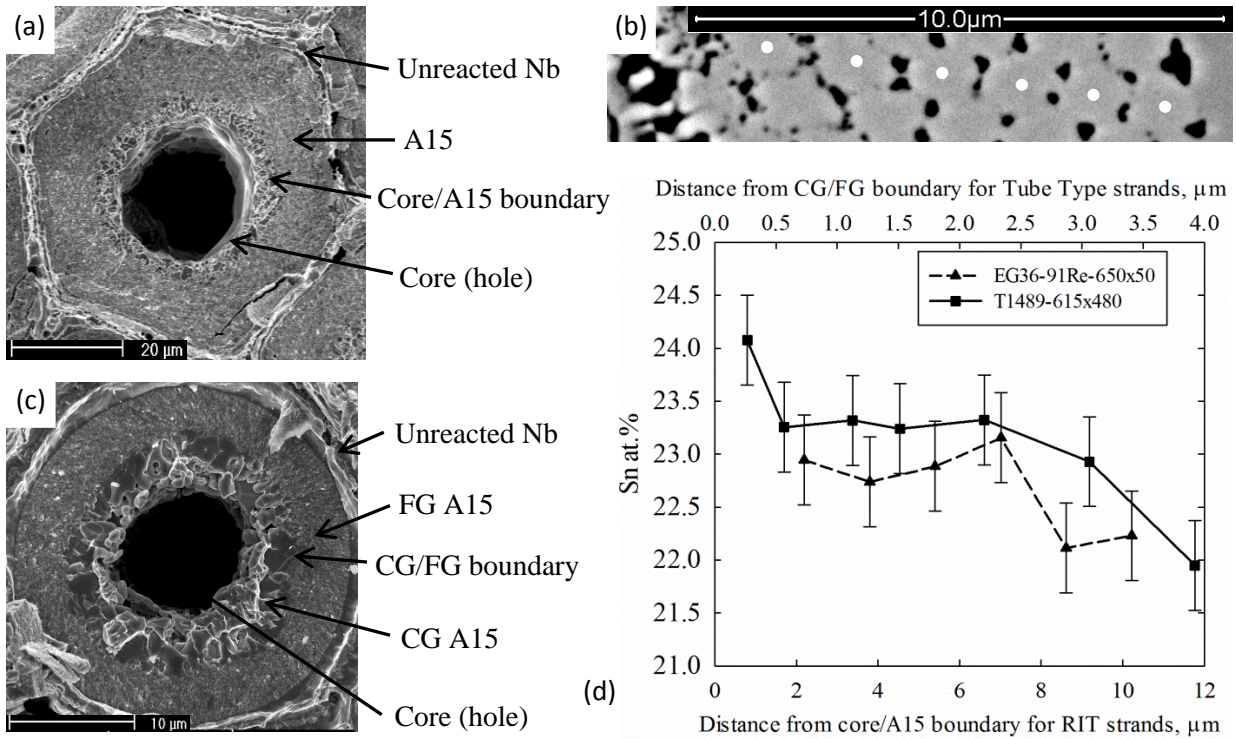

Figure 4. (a) The fracture SEM image of a subelement of EG36-91Re-650x50 showing the core and A15 regions; (b) The BSE/SEM image of EG36-91Re-650x50, with the white dots indicating the spots where EDS were performed; (c) the fracture SEM image of a subelement of T1489-615x480 showing the CG and FG A15 regions; (d) the Sn concentration profiles of EG36-91Re-650x50 and T1489-615x480h strands.

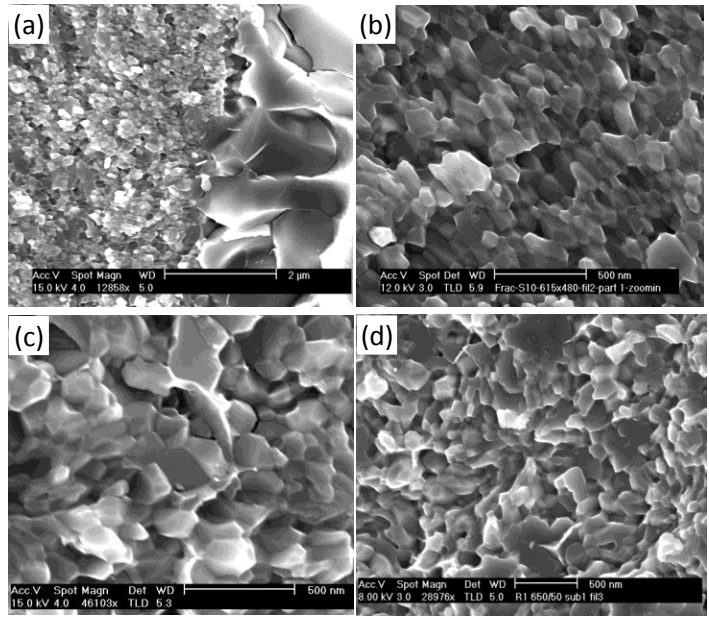

Figure 5. Fracture SEM images of (a) FG and CG area of T1489-615x480, (b) FG area of T1489-615x480, (c) A15 area of EG36Ti-61Re-650x60, (d) A15 area of EG36-91Re-650x50.

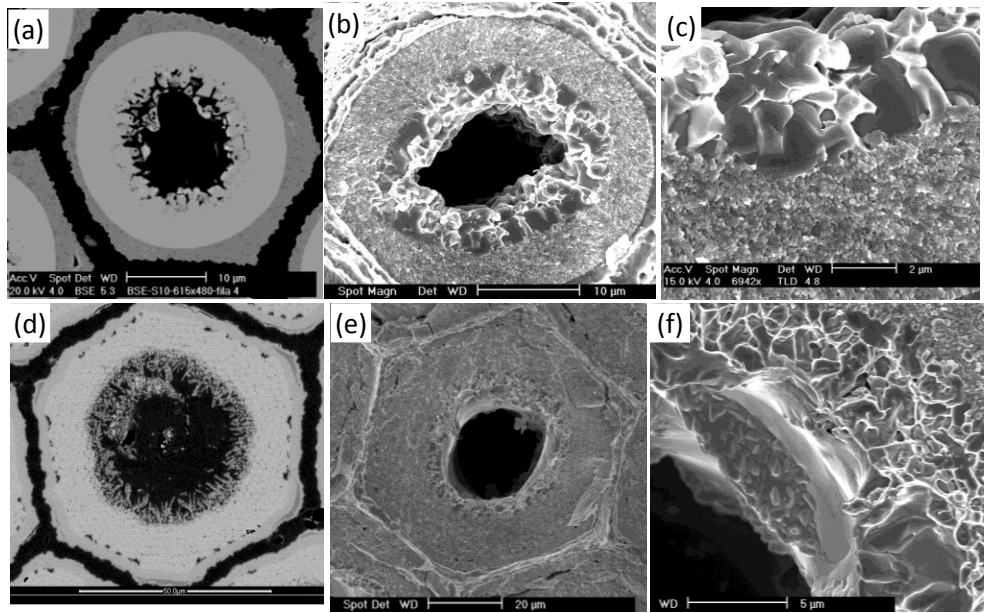

Figure 6. SEM images of Tube Type and RIT strands: (a) BSE image of T1489-615x480, (b) fracture SEM of T1505-625x500, (c) fracture SEM of T1505-625x500 showing the core/A15 interface (d) BSE image of HP31-61Re-650x80, (e) fracture SEM of EG36Ti-61Re-650x60, (f) fracture SEM of EG36Ti-61Re-650x60 showing the core/A15 interface.